\documentclass{article}
\usepackage{arxiv}
\usepackage{amssymb}
\usepackage[utf8]{inputenc} 
\usepackage[T1]{fontenc}    
\usepackage{url}   		 
\usepackage{booktabs}  	 
\usepackage{amsfonts}  	 
\usepackage{nicefrac}  	 
\usepackage{microtype} 	 
\usepackage{lipsum}  	 
\usepackage{graphicx}  	
\usepackage{subcaption}
\usepackage{doi}
\usepackage{amsmath, calc}
\usepackage{float}
\usepackage{mathabx}
\usepackage{apacite}
\usepackage[toc,page]{appendix}
\usepackage{listings}
\usepackage{xcolor}
\usepackage{setspace} 
\usepackage{natbib}
\usepackage{multicol}
\usepackage{multirow}
\usepackage{array}
\usepackage{longtable}
\usepackage{colortbl}
\definecolor{lightgray}{gray}{0.7}  
\newcommand{\z}{\textcolor{lightgray}{0}}
\newcommand{\bic}{$\text{BIC}_{m}$\:}
\newcommand{\waic}{$\text{WAIC}_{c}$\:}

\definecolor{codegreen}{rgb}{0,0.6,0}
\definecolor{codegray}{rgb}{0.5,0.5,0.5}
\definecolor{codepurple}{rgb}{0.58,0,0.82}
\definecolor{backcolour}{rgb}{0.95,0.95,0.92}

\lstdefinestyle{mystyle}{
    backgroundcolor=\color{backcolour},   
    commentstyle=\color{codegreen},
    keywordstyle=\color{magenta},
    numberstyle=\tiny\color{codegray},
    stringstyle=\color{codepurple},
    basicstyle=\ttfamily\footnotesize,
    breakatwhitespace=false,   	 
    breaklines=true,       		 
    captionpos=b,          		 
    keepspaces=true,       		 
    numbers=left,          		 
    numbersep=5pt,        		 
    showspaces=false,      		 
    showstringspaces=false,
    showtabs=false,        		 
    tabsize=2
}

\title{Summarising mortality data with a time-dependent beta latent variable model}

\date{}   				  

\author{ Pedro Menezes de Araújo\\
	School of Mathematics and Statistics\\
	University College Dublin\\
	\texttt{pedro.menezesdearaujo@ucdconnect.ie} \\
	\And
	Isobel Claire Gormley \\
	School of Mathematics and Statistics\\
	University College Dublin\\
	\texttt{claire.gormley@ucd.ie} \\
	\And
	Thomas Brendan Murphy \\
	School of Mathematics and Statistics\\
	University College Dublin\\
	\texttt{brendan.murphy@ucd.ie} \\
}

\renewcommand{\headeright}{}
\renewcommand{\undertitle}{}
\renewcommand{\shorttitle}{Time-dependent BLV}

\hypersetup{
pdftitle={Summarising mortality data with a time-dependent beta latent variable model}
}

\begin{document}

\maketitle

\begin{abstract}

Age-specific probabilities of death provide a snapshot of population mortality at the country level at a given point in time. Due to the high dimensionality of the data, summarising mortality information is essential for various analyses, such as visualisation and clustering. We propose the use of beta latent variable (BLV) models to summarise mortality information without data transformation. A time-dependent version of the BLV model is developed by incorporating an autoregressive prior for the latent effects. This model aims to represent mortality data with a small set of $K$ latent effects while accounting for time dependence between these effects. Inference is performed using Bayesian methods, with posterior samples generated via Hamiltonian Monte Carlo. The BLV model is applied to probabilities of death from the Human Mortality Database, covering 41 countries and 23 age-specific probabilities of death over several periods. The time-dependent BLV model with $K=6$ latent effects accurately reconstructs observed mortality data, and the model parameters have intuitive and insightful interpretations. The time-dependent BLV outperforms the standard Gaussian factor analysis model applied to logit probability of death, and demonstrates that BLV models can effectively summarise mortality data. 

\end{abstract}


\section{Introduction}

Probabilities of death quantify the mortality levels of a population and are crucial for guiding public and private institutions concerned with, for instance, public health and insurance. Data are usually recorded across age groups, time, and sex, and are available for several countries, resulting in a complex multivariate representation of countries' mortality profiles. 
Given the importance and complexity of mortality data, appropriate statistical analysis is essential. Several models have been used to analyse mortality data, with the Lee-Carter model \citep{Lee_Carter_1992} and its extensions among the leading methods for forecasting. Beyond forecasting, there is interest in unsupervised approaches aimed at summarising and understanding countries' mortality profiles. Studies such as \cite{Debon_Chaves_2017}, \cite{Carracedo_Debon_2018}, \cite{Fop_Murphy_2018}, \cite{Leger_Mazzuco_2021}, and \cite{Tsai_Chi-Liang_2021} focus on clustering country-level mortality data, while \cite{Fosdick_Hoff_2014} models mortality using a factor model to handle missing data imputation. In many studies, especially those related to multi-population models, summarising mortality data is an important step due to the high dimensionality of the data, which makes modelling and visualisation challenging. A common approach consists of first transforming mortality data, then applying a dimension reduction technique, and finally employing methods for tasks related to life table estimation, forecasting, clustering, etc. For instance, principal component analysis (PCA) can be applied to the probability of death to estimate model life tables \citep{UN_1982}; the Lee-Carter model consists of applying singular value decomposition to the log central mortality rates before forecasting it; \cite{Mesle_Vallin_2002} applied principal component analysis (PCA) to the log probability of death for clustering purposes. Similarly, \cite{Debon_Chaves_2017} applied PCA to the logit probability of death before clustering, while \cite{Leger_Mazzuco_2021} used functional principal component analysis to better visualise mortality data. 

In this work, we propose using beta latent variable models to summarise mortality data more effectively. Logit or log transformations are commonly applied due to the absence of suitable models for bounded data over time. However, such transformations can distort key data features \citep{Revuelta_Hidalgo_2020} and, as implied by Jensen's inequality, may lead to biased predictions when transforming estimates back to the original scale.

We propose a country-specific, time-dependent beta latent variable model (BLV) to model and summarise mortality data effectively. Beta latent variable models are related to beta factor models \citep{Revuelta_Hidalgo_2020} and beta item response theory models \citep{Noel_Dauvier_2007}. The BLV model serves as an alternative to the Gaussian factor model \citep{Garson_2022}. We extend the classical BLV model by incorporating an autoregressive prior for the latent effects, accommodating temporal associations. This approach allows us to explain observed mortality using a small set of latent, time-specific effects, thereby summarising the data and enhancing visualisation to improve understanding of countries' mortality dynamics.

The time-dependent BLV model is estimated using Bayesian inference, with samples drawn from the posterior distribution via Hamiltonian Monte Carlo (HMC) \citep{Betancourt_2018}, implemented through the Stan programming language \citep{Stan_2024} in \texttt{R} \citep{R_2021}. To address model non-identifiability issues, samples are post-processed using Procrustean matching \citep{Borg_Groenen_2005}.

For model selection, we propose a Monte Carlo approximation of the Bayesian information criterion (BIC) based on the marginal likelihood with respect to the latent effects. The widely applicable information criterion (WAIC), based on the conditional likelihood \citep{Gelman_Hwang_2013, Luo_Al-Harbi_2017}, is also considered. As is standard statistical practice, model fit is assessed using posterior predictive distributions and compared with a baseline based on the logit probability of death.

We explore mortality data from the Human Mortality Database (HMD), which compiles and distributes estimates of probabilities of death and related statistics for various populations (mainly developed countries) \citep{HMD}. The data are publicly available at \href{https://www.mortality.org/}{www.mortality.org} and have been analysed by works such as \cite{Clark_Sharrow_2011}, \cite{Debon_Chaves_2017}, \cite{Leger_Mazzuco_2021},  \cite{Shen_Li_2024} and many others.

In what follows, Section \ref{section:data} describes the HMD data and provides preliminary analyses. Section \ref{section:model} introduces the BLV model and discusses prior distributions, inference, model selection, and model fit. We present the results of modelling the HMD data using the time-dependent BLV model in Section \ref{section:results}, and conclude in Section \ref{section:conclusion} with a discussion of future work. The Stan code for fitting the developed BLV model is available at \href{https://github.com/pedroaraujo9/btblv}{https://github.com/pedroaraujo9/btblv} and the code for the data analysis and simulation study is available at \href{https://github.com/pedroaraujo9/tblv}{https://github.com/pedroaraujo9/tblv}.

\section{The Human Mortality Database}\label{section:data}

We use data from the Human Mortality Database (HMD), available at \href{https://www.mortality.org/}{www.mortality.org}. The HMD is an open data resource that compiles mortality tables and other statistics related to mortality from several countries \citep{HMD}. Data can be retrieved through the \texttt{R} package \texttt{HMDHFDplus} \citep{Riffe_2015}.

\subsection{Mortality data}

We consider 5-year probability of death from period life tables aggregated for age groups with a 5-year-age-size $[x, x+5)$, which is typically defined as \citep{Preston_Heuveline_2000,Dodd_Foster_2017}:
$${}_{5}q_x=\frac{{}_{5}d_x}{{}_{5}l_x}=\frac{\text{Deaths between ages } x\text{ and }x+5}{\text{Number of survivors between ages } x \text{ and }x+5}.$$  

Here, combined (i.e., non-gender specific) age group probabilities of death were considered for 5-year-size periods. The first age group interval is $[0, 1)$ years, with subsequent intervals $[1, 5), [5, 10), [10, 15), \dots, [105, 110)$. Throughout, we denote an age group interval by the first value $x$ of the interval, hence $ x\in \mathcal{X}=\{0, 1, 5, 10, 15, \dots, 105\}$, resulting in $23$ probabilities of death for each country at each time point. While different countries have varying time series lengths (as detailed in Appendix \ref{appendix:data-availability}), no country has missing values within its time series.

The periods analysed are from 1950 to the end of 2019, with a 5-year time interval length. We denote each time point by its order, where the first interval $[1950, 1955)$ is $t=1$, and the last $[2015,2020)$ is $t=14$. For clarity, we suppress the left subscript 5 and denote the probabilities of death $_{5}q_{x}$ by $q_{xit}$ representing the probability of death for the age group $x$, country $i$ at period $t$. Some countries have fewer data available, and $s_i$ denotes the first time point for country $i$ and $T_i$ as the last time point available for country $i$; additionally, the total number of time points for country $i$ is denoted $N_i$. The availability of data in countries over time is detailed in Appendix \ref{appendix:data-availability}.

In summary, we have observed probabilities of death for 23 age groups in $n=41$ countries over time, with different numbers of time points for each country. Our final data $\mathbf{q}$ is a $484\times 23$ matrix defined as a collection of row vectors $\mathbf{q}_{it} = (q_{0it}, \dots, q_{105, it})$ with the age-specific probabilities of death (our variables) for each country $i$ at time $t$.

\subsection{Mortality trends and associations}

\begin{figure}[htbp]
  \centering
  \begin{tabular}{@{}cc@{}} 
    \includegraphics[scale=0.62]{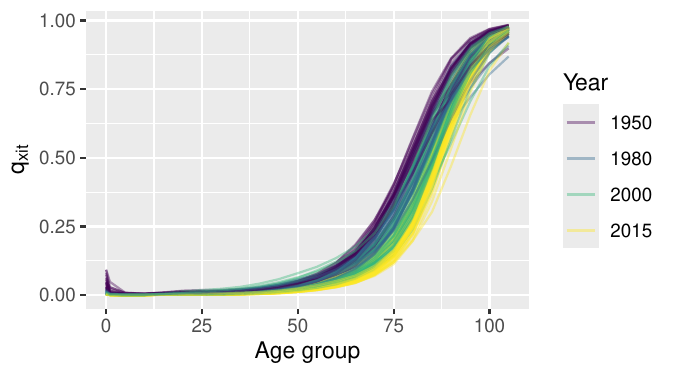} &
    \includegraphics[scale=0.62]{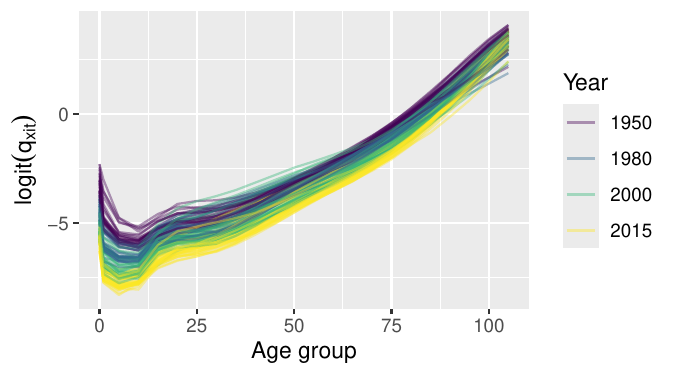} \\
    \small (a) Probabilities of death on the original scale. & \small (b) Probabilities of death on the logit scale.
  \end{tabular}
  \caption{Probabilities of death on the original (a) and logit (b) scale for all countries for some selected periods.}\label{fig:data:mort_curves}
\end{figure}

Figure \ref{fig:data:mort_curves} illustrates the mortality curves $\mathbf{q}_{it}$ in the original and logit scales for all countries for $i=1, \ldots, n$ at a subset of periods, for visual clarity. Temporal effects are apparent; for instance, lower mortality is observed for younger age groups in the recent periods (2000 and 2015), and higher mortality is observed for the countries in the first period (1950) analysed. There is also a difference in the scale of the mortality, with very low values for younger age groups and high mortality for older age groups.

Kendall's $\tau$ coefficient \citep{Millard_2013} can be used to quantify the probability of death trend. For the probability of death series at an age group $x$ and country $i$, $(q_{xi1}, \dots, q_{xiT})$, it is defined as
$$\tau = \frac{2S}{T(T-1)}, \:\text{where } S = \sum_{t=1}^{T-1}\sum_{t'=t+1}^{T}\text{sign}\{(q_{xit}-q_{xit'})\}.$$

\begin{figure}[H]
\centering
\includegraphics[width=0.85\textwidth]{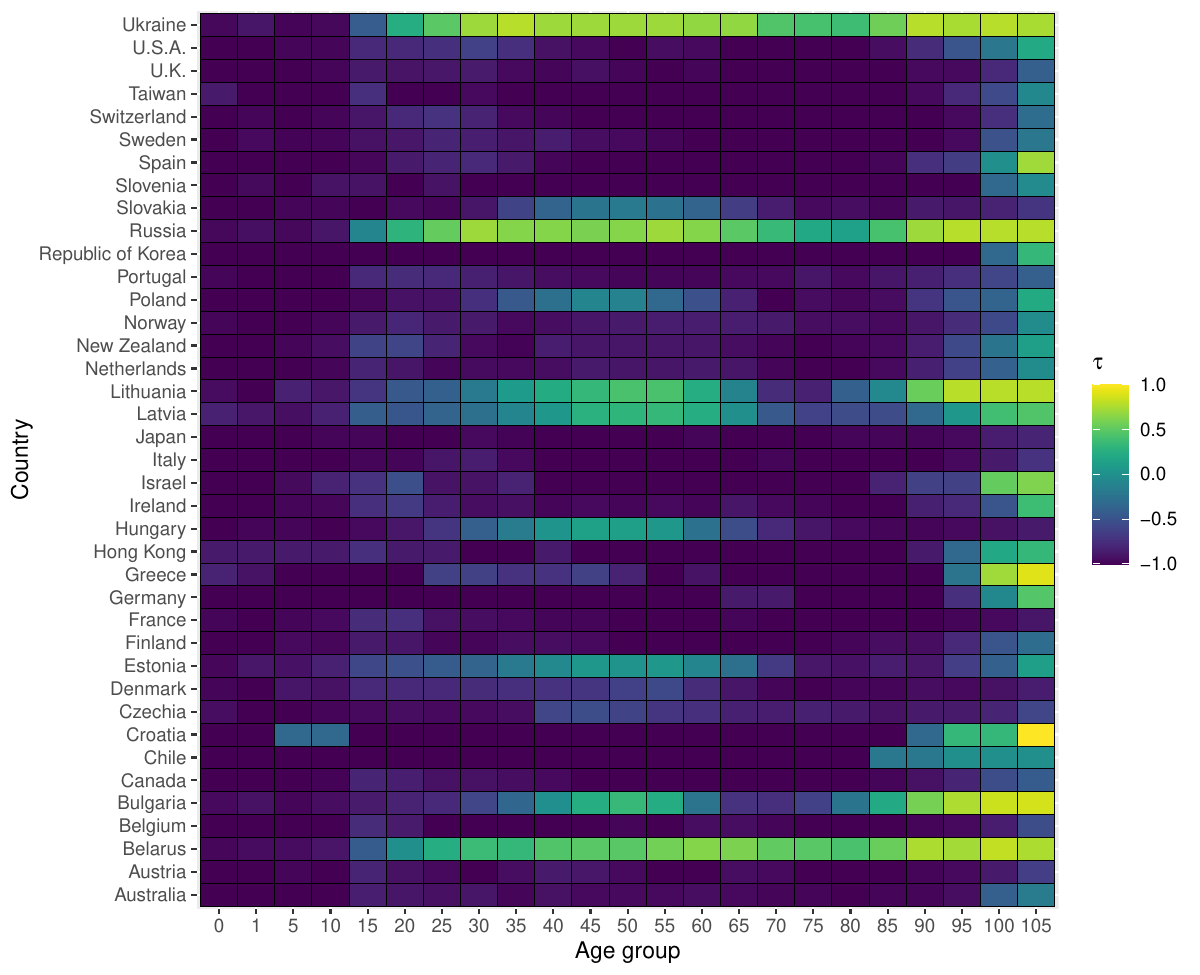}
\caption{Estimated Kendall's $\tau$ coefficient for the probability of death trend for different age groups and countries.}
\label{fig:data:tau}
\end{figure}

The coefficient lies in the interval $[-1, 1]$, with $\tau = 0$ meaning there is no trend in the data, $\tau = 1$ meaning there is a positive monotonic trend, and $\tau = -1$ meaning there is a negative monotonic trend. Figure \ref{fig:data:tau} shows $\tau$ for all countries and age groups. We see that probabilities of death have decreased in all countries for younger age groups such as $x \in \{0, 1, 5, 10\}$. For some countries, such as Russia, Belarus, and Ukraine, we see a weak positive probability of death trend from mid-age to older age groups, indicating more variability in the probability of death trajectory. Generally, the probability of death trend is near zero or greater than zero for the older age groups such as $x\geq 100$.

\begin{figure}[H]
\centering \includegraphics[scale=0.7]{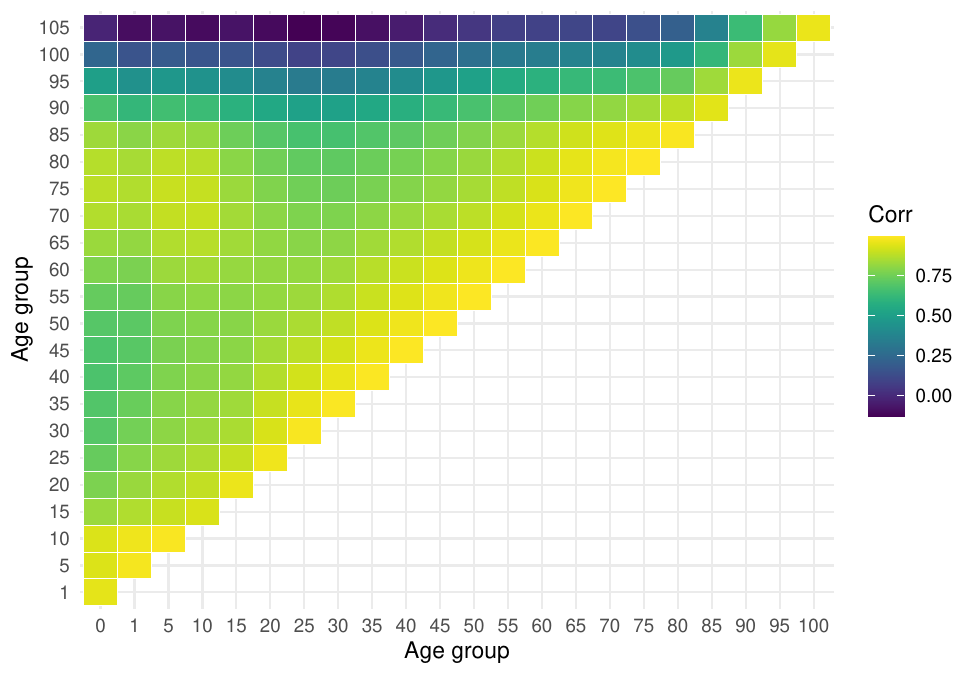} \caption{Correlation matrix of age group mortalities, calculated using all country–year pairs as observations and age-group probabilities of death as variables.} \label{fig:data:corr} \end{figure}

Figure \ref{fig:data:corr} shows the Pearson correlations between the age group probabilities of death (the data matrix $\mathbf{q}$). We see some blocks of contiguous age groups with very high correlations, for example, the age intervals $x \in \{0, 1, 5, 10\}$, and $x \in \{65, \dots, 85\}$. These blocks suggest that mortality in some age groups is possibly associated with a common latent effect, suggesting that we can summarise the mortality profile with a small set of latent effects.

 \section{A Bayesian time-dependent beta latent variable model}\label{section:model}

\subsection{The beta distribution}

Most analyses of $(0,1)$-bounded data rely on transformations such as the logarithm or logit because suitable models for the raw scale are limited. Such transformations can alter key features of the data \citep{Revuelta_Hidalgo_2020} and affect the estimation of expected mortality on the original scale (via Jensen's inequality). Probabilities of death are bounded between 0 and 1, and the beta distribution is a natural choice for modelling such data. 

We consider the beta probability distribution parameterised in terms of its expected value $\mu$ and precision parameter $\kappa$, and refer to it as the beta-proportion distribution denoted by $\text{Beta-prop}(\mu, \kappa)$, where $\mu \in (0, 1)$ and $\kappa \in \mathbb{R}^{+}$. This distribution is widely used in regression problems \citep{Ferrari_Cribari-Neto_2004} for modelling probabilities and rates and has the following probability density function
$$p(y\mid \mu, \kappa) = \frac{y^{\kappa\mu-1}(1-y)^{\kappa(1-\mu)-1}}{B(\kappa\mu, \kappa(1-\mu))},$$
where $B(\cdot, \cdot)$ is the beta function. The more typical parameterisation of the beta distribution in terms of the two shape parameters $a, b$ is obtained by setting $a = \kappa\mu$ and $b = \kappa(1-\mu)$. 

Here we assume $q_{xit}\mid \mu_{xit}, \kappa\sim \text{Beta-prop}(\mu_{xit}, \kappa)$, therefore, $\text{E}(q_{xit}\mid \mu_{xit}, \kappa) = \mu_{xit}, \text{ and }\text{Var}(q_{xit}\mid \mu_{xit}, \kappa) = \frac{\mu_{xit}(1-\mu_{xit})}{1+\kappa}$ where $\mu_{xit}$ represents the expected mortality for age group $x$, country $i$, and time $t$, and $\kappa$ is a common precision parameter. 

\subsection{A beta latent variable model}

As observed in Figure \ref{fig:data:corr}, the mortality data are highly correlated, indicating that a latent structure may assist in explaining the associations between the probabilities of death in some age groups. We estimate these latent effects by modelling the logit of the expected mortality with a linear beta latent variable model that accounts for the bounded nature of the data, i.e.,

\begin{equation}\label{model-structure}
    \text{logit}(\mu_{xit}) = \log\bigg(\frac{\mu_{xit}}{1-\mu_{xit}}\bigg) = \beta_{x} + \sum_{k=1}^K \alpha_{xk}\theta_{ik}^{(t)} =  \beta_{x} + \boldsymbol{\alpha}_{x}^{'}\boldsymbol{\theta}_{i}^{(t)},
\end{equation}
where $\boldsymbol{\theta}_{i}^{(t)} = (\theta_{i1}^{(t)},\dots,\theta_{iK}^{(t)})^{'}$ and $\boldsymbol{\alpha}_{x}=(\alpha_{x1},\dots,\alpha_{xK})^{'}$ are $K$ dimensional vectors, where $K$ is the unknown number of latent effects. The expected mortality is a function of the inverse of the logit function, i.e.:

\begin{equation}\label{equation:BLV-mu}
	\mu_{xit} = \frac{\exp(\beta_{x} + \boldsymbol{\alpha}_{x}^{'}\boldsymbol{\theta}_{i}^{(t)})}{1+\exp(\beta_{x} + \boldsymbol{\alpha}_{x}^{'}\boldsymbol{\theta}_{i}^{(t)})} = \eta(\boldsymbol{\theta}_{i}^{(t)}, \boldsymbol{\alpha}_{x}, \beta_{x}).
\end{equation}

While the likelihood models the observed mortality data directly on their original scale, the logit transformation is applied only to the mean parameter $\mu_{xit}$. This ensures that the latent structure maps to the unit interval and is standard in generalised latent variable models.

Regarding the parameters, $\beta_{x}$ represents the baseline level of (logit) mortality of the age group $x$, and is invariant over time since we want to capture the effect of the country on mortality when compared to the baseline age-specific mortality. The parameter $\alpha_{xk}$ is the loading or coefficient related to age group $x$ and latent dimension $k$ and represents the strength of the relationship between age group $x$ and latent effect dimension $k$. For instance, $\alpha_{xk} = 0$ indicates that the age group $x$ is unrelated to the latent effect dimension $k$.

The latent effect for country $i$, dimension $k$ and time $t$ is denoted $\theta_{ik}^{(t)}$ and we define the vector $\boldsymbol{\theta}_{i}^{(t)}=(\theta_{i1}^{(t)},\dots, \theta_{iK}^{(t)})^{'}$ as the latent effects for country $i$ at time $t$. The latent effects represent how much lower or higher the probability of death in country $i$ at time $t$ is when compared to the baseline mortality $\beta_{x}$ of age group $x$, scaled by the associated $\alpha_{xk}$. 
In most applications, $\theta_{ik}^{(t)}\overset{iid}{\sim} N(0, \sigma^{2})$, like in \cite{Revuelta_Hidalgo_2020} and \cite{Noel_Dauvier_2007}. However, to account for the time dependence for each country, we propose a group-specific time-dependent distribution for the latent effects, as discussed in Section \ref{theta-prior}.

For the precision parameter, we opt for parsimony and constrain all age groups to have the same precision $\kappa$. The mortality variance is, therefore, $\mu_{xit}(1-\mu_{xit})(1+\kappa)^{-1}$. This implies that the variance depends mainly on the mean, scaled by a common $\kappa$, imposing the restriction that the noise around the latent-effect–induced mean is similar across age groups with a similar scale. While specifying the model with age group specific precision parameters is feasible and would increase model flexibility, in practise such models were prone to overfitting.

In summary, the beta latent variable (BLV) model defined in (\ref{model-structure}) has a standard linear form, similar to latent variable models for other distributions \citep{Skrondal_Rabe-Hesketh_2004} and has a similar structure to beta item response theory models \citep{Noel_Dauvier_2007} and beta factor models \citep{Revuelta_Hidalgo_2020}. By considering $\text{logit}(q_{xit})$ with a Gaussian distribution and using the same structure as in (\ref{model-structure}), we have the well-known Gaussian factor analysis model \citep{Garson_2022}. The Gaussian case, when $K = 1$, is similar to the Lee-Carter model \citep{Lee_Carter_1992}, but it is not country-specific, being a more general linear latent representation of the logit-mortality.

\subsection{A country-specific time-dependent prior distribution for the latent effects}\label{theta-prior}

Figures \ref{fig:data:mort_curves} and \ref{fig:data:tau} suggest that there is variability in the mortality trends across countries and that the probabilities of death are associated over time. Therefore, it is reasonable to assume that the country's latent effects are dependent over time rather than assuming independence, as is typically the case for BLV models. An AR(1) structure is assumed for the latent effects, where $\theta_{ik}^{(t)}\sim N(\phi_{i}\theta_{ik}^{(t-1)}, \sigma_{i}^2)$ where $\phi_{i}\in (-1, 1)$ to ensure stationarity, and for the initial time point $\theta_{ik}^{(s_i)}\sim N(0, \frac{\sigma_{i}^2}{1-\phi_{i}^2})$, which is the marginal distribution of the AR(1) process. For the AR(1) parameters, we assume $\log(\sigma_{i})\overset{iid}{\sim} \text{Normal}(0, 1)$ and $\phi_{i}\overset{iid}{\sim} U(-1, 1)$, where $U$ is the uniform distribution over the given interval.

The country-specific, time-dependent distribution for the latent effects in vector form, with a non-central representation, is therefore
\begin{equation}\label{equation:prior-theta}
	\boldsymbol{\theta}_{i}^{(t)} = \phi_{i}\boldsymbol{\theta}_{i}^{(t-1)} + \sigma_{i}\boldsymbol{\epsilon}_{i}^{(t)}  =  \sigma_{i}\frac{\phi_i^{t-1}}{\sqrt{1-\phi_i^2}}  \   \boldsymbol{\epsilon}_{i}^{(s_i)} +  \sigma_{i}\sum_{j=0}^{t-2} \phi_{i}^{j}\boldsymbol{\epsilon}_{i}^{(t-j)}\, \text{\:\:for\:} t > s_i,
\end{equation}
where $\boldsymbol{\epsilon}_{i}^{(t)} = (\epsilon_{i1}^{(t)}, \dots, \epsilon_{iK}^{(t)})^{'}$ follows a standard multivariate normal distribution i.e., $\boldsymbol{\epsilon}_{i}^{(t)} \sim \text{MVN}(\mathbf{0}_{K}, \mathbf{I}_{K})$. This non-central representation is more stable when using Monte Carlo sampling; therefore, when estimating the parameters or approximating quantities, we treat $\boldsymbol{\epsilon}_{i}^{(t)}$ as the latent variable and transform it back to $\boldsymbol{\theta}_{i}^{(t)}$ as required.

The main goal of this prior distribution is to summarise and describe the time dependence and variability within each country in a simple way. The parameter $\phi_i$ is related to the autocorrelation $\text{Corr}(\theta_{ik}^{(t)}, \theta_{ik}^{(s)}) = \phi_{i}^{|t-s|}$, and $\sigma_{i}$ to the scale of the latent effects. Therefore, $\phi_i$ and $\sigma_i$ can be seen as an extra tool to summarise and compare the countries' latent effects given that countries with similar $\phi_i$, and $\sigma_i$ will have a similar covariance matrix for $\boldsymbol{\theta}_{i}^{(t)}$.

\subsection{The time-dependent beta latent variable model}\label{prior-age-group}

For the BLV model coefficients, we assume $\alpha_{xk}\sim N(0, 1)$, a common choice for the prior of coefficients for latent variable models \citep{Bazan_Bolfarine_2006,Murphy_Viroli_2020}. For the parameters $\beta_{x}$, we assume a vague prior with $\beta_{x}\sim N(0, 100)$ and for the precision parameter, we assume a prior $\log(\kappa)\sim N(0, 100)$. The hierarchical representation for the country-specific, time-dependent BLV model, given $\eta(\boldsymbol{\theta}_{i}^{(t)}, \boldsymbol{\alpha}_{x}, \beta_{x})$, as defined in (\ref{equation:BLV-mu}), is then:

\begin{align*}
	q_{xit}\mid\boldsymbol{\theta}_{i}^{(t)}, \boldsymbol{\alpha}_{x}, \beta_{x}, \kappa &\sim \text{Beta-prop}\{\eta(\boldsymbol{\theta}_{i}^{(t)}, \boldsymbol{\alpha}_{x}, \beta_{x}), \kappa\}, \\
	\boldsymbol{\theta}_{i}^{(t)}& = \phi_{i}\boldsymbol{\theta}_{i}^{(t-1)} + \sigma_{i}\boldsymbol{\epsilon}_{i}^{(t)}, \\
	\boldsymbol{\epsilon}_{i}^{(t)} &\sim \text{MVN}(\mathbf{0}_{K}, \mathbf{I}_{K}), \\
	\phi_{i} &\sim U(-1, 1), \\
	\log(\sigma_{i}) &\sim N(0, 1), \\
   \alpha_{xk} &\sim N(0, 1), \\
   \beta_{x} &\sim N(0, 100), \\
	\log(\kappa) &\sim N(0, 100). \\
\end{align*}

\subsection{Model inference}

For a general mortality modelling problem with age groups $x\in\mathcal{X}$ (with $J$ levels), countries $i = 1,\dots, n$, and time points $t = s_i,\dots, T_i$ for each country $i$, let $\boldsymbol{\theta}=(\boldsymbol{\theta}_{1}^{(s_1)},\dots, \boldsymbol{\theta}_{n}^{(T_n)})$ denote all latent effects, $\Lambda = (\boldsymbol{\alpha}, \boldsymbol{\beta}, \kappa)$ denote all age group parameters where $\boldsymbol{\alpha} = (\boldsymbol{\alpha}_{0},\dots,\boldsymbol{\alpha}_{J})$ with $\boldsymbol{\alpha}_x = (\alpha_{x1},\dots,\alpha_{xK})$, $\boldsymbol{\beta}=(\beta_{0}, \dots, \beta_{J})$, and let $\boldsymbol{\phi}=(\phi_{1}, \dots, \phi_{n})$ and $\boldsymbol{\sigma}=(\sigma_{1},\dots,\sigma_{n})$ denote the country-specific autoregressive parameters. The BLV model conditional likelihood function is

\begin{equation}\label{equation:cond-likelihood}
	p(\mathbf{q}\mid \boldsymbol{\epsilon}, \Lambda, \boldsymbol{\phi}, \boldsymbol{\sigma}) = \prod_{i=1}^{n}\prod_{t=s_i}^{T_i}\prod_{x\in\mathcal{X}}\frac{q_{xit}^{\kappa\mu_{xit}-1}(1-q_{xit})^{\kappa(1-\mu_{xit})-1}}{B(\kappa\mu_{xit}, \kappa(1-\mu_{xit}))},
\end{equation}
where $\mu_{xit} = \eta(\boldsymbol{\theta}_{i}^{(t)}, \boldsymbol{\alpha}_{x}, \beta_{x})$ as in (\ref{equation:BLV-mu}), and $\boldsymbol{\theta}_{i}^{(t)}$ is a function of the increments $\boldsymbol{\epsilon}_{i}^{(t)}$ as defined in (\ref{equation:prior-theta}). A key assumption here is the conditional independence between the observed probabilities of death, given the model parameters and latent effects, which allows factorisation of the joint density into a product.

The marginal likelihood of the BLV model is the conditional likelihood (\ref{equation:cond-likelihood}) with the latent effects integrated out, meaning countries remain conditionally independent, but the probabilities of death observed within a country are no longer independent. Denoting the multiple required integrals as a single integral and $\boldsymbol{\epsilon}_{i}$ as the collection of all country latent effects to simplify the notation, the BLV marginal likelihood is

\begin{equation}\label{equation:marg-likelihood}
	p(\mathbf{q}\mid\Lambda, \boldsymbol{\phi}, \boldsymbol{\sigma}) = \prod_{i=1}^{n}\int_{-\infty}^{\infty}\bigg\{\prod_{t=s_i}^{T_i}\prod_{x\in\mathcal{X}}\frac{q_{xit}^{\kappa\mu_{xit}-1}(1-q_{xit})^{\kappa (1-\mu_{xit})-1}}{B(\kappa\mu_{xit}, \kappa(1-\mu_{xit}))}\prod_{t=s_i}^{T_i}p(\boldsymbol{\epsilon_i}^{(t)})\bigg\}d\boldsymbol{\epsilon}_{i}.
\end{equation}

The integral in (\ref{equation:marg-likelihood}) is complex to evaluate, which makes, e.g., maximum likelihood inference difficult. However, Bayesian methods provide an alternative approach to estimate the parameters of latent variable models since they use the conditional likelihood (\ref{equation:cond-likelihood}) and can deal well with non-identifiability issues.

Given the priors for the model parameters defined in Section \ref{theta-prior}, the posterior distribution of the time-dependent BLV model has no closed form; therefore, we generate samples from the posterior using Hamiltonian Monte Carlo (HMC). Hamiltonian Monte Carlo methods require the existence of the gradient of the log-posterior distribution to update the Markov Chain, finding an optimal proposal for the chain based on Hamiltonian dynamics. This proposal can lead to very weakly autocorrelated chains and faster convergence than traditional Markov chain Monte Carlo (MCMC) random-walk-based methods, such as Metropolis-Hastings. More details about HMC can be found in \cite{Homan_Gelman_2014} and \cite{Betancourt_2018}. For starting points for the chains, we sample from $U(-0.1, 0.1)$ for $\phi_{i}$; from $N(0, 0.1^2)$ for $\alpha_{xk}$ and $\theta_{ik}^{(t)}$ to avoid extreme values; and for $\beta_{x}$ and $\log(\kappa)$ we sample from $N(\beta_{x, \text{MLE}}, 0.1^2)$ and $N(\log{(\kappa)}_{\text{MLE}}, 0.1^2)$ respectively, where the MLE subscript denotes the MLE estimate for the BLV model where $\alpha_{xk} = 0$ for all $x$ and $k$, which we fit with the \texttt{R} package \texttt{betareg} \citep{Cribari-Neto_Zeileis_2010}.

We use the \texttt{Stan} programming language to implement the HMC algorithm via the R package \texttt{rstan} \citep{Stan_2024}; we use the algorithm's NUTS HMC \citep{Homan_Gelman_2014}, which finds optimal values for the HMC hyperparameters. 

The use of the non-central representation for the time-dependent latent effects (\ref{equation:prior-theta}) improves the computational efficiency of the HMC algorithm. The marginal posterior mean is used as the Bayes estimator for all the model parameters, which we approximate with the sample generated by the HMC algorithm. Highest posterior density (HPD) intervals are also available for the model parameters; we consider $95\%$ HPDs, which provide the widest interval where the probability of the parameter being in it is 0.95.

\subsubsection{Identifiability}

The time-dependent BLV model is not identifiable with respect to the latent effect label $k$ nor for rotations and translations of $\boldsymbol{\alpha}_{x}$ and $\boldsymbol{\theta}_{i}^{(t)}$. This means that at each iteration of the HMC, we may have different values for the parameters leading to the same value of the posterior density. We solve this problem in an offline manner, using Procrustean matching \citep{Borg_Groenen_2005}. Specifically, we select as a reference $\boldsymbol{\alpha}_{x}^{*}$, the loadings from the application of principal component analysis (PCA) to the correlation matrix of the $\text{logit}(q_{xit})$ data to have a standard reference. We then find the orthogonal Procrustean transform $K\times K$ matrix $\mathbf{L}^{(j)}$ for each iteration $j$ that minimises the squared error between $\boldsymbol{\alpha}_{x}^{(j)}$ at iteration $j$ and the reference $\boldsymbol{\alpha}_{x}^{*}$.

\subsection{Model choice}

To complete the model definition, the number of latent effects $K$ must be specified. We expect $K \ll |\mathcal{X}|$, such that a few latent effects would capture almost all the information in the $|\mathcal{X}|$ age group probabilities of death. To select $K$, the use of the marginal likelihood is desirable, as this would ensure the generalisability of the time-dependent BLV to out-of-sample observations \citep{Merkle_Edgar_2019}.

The time-dependent BLV model marginal likelihood (\ref{equation:marg-likelihood}) is not readily available. Given that we are assuming conditional independence between the countries, we can approximate the marginal likelihood for each country with Monte Carlo integration. We use importance sampling \citep{Robert_Casella_2009} to reduce the variance of the estimator, where the proposal distribution $g\sim \text{MVN}(\hat{\boldsymbol{\epsilon}}_i^{(t)}, \hat{\Sigma}_{it})$ is the multivariate normal distribution centred at the posterior mean vector $\hat{\boldsymbol{\epsilon}}_i^{(t)}$, with the covariance matrix $\hat{\Sigma}_{it}$ being the posterior covariance matrix of $\boldsymbol{\epsilon}_i^{(t)}$. The proposal $g$ is a normal approximation of the posterior distribution of the vector $\boldsymbol{\epsilon}_i^{(t)}$. In summary, the marginal likelihood is approximated based on the expression:

\begin{equation}\label{equation:country-marginal}
	\log p(\mathbf{q}\mid\Lambda, \boldsymbol{\phi}, \boldsymbol{\sigma}) = \sum_{i=1}^{n}\log\text{E}_{g}\bigg\{\prod_{t=s_i}^{T_i}\prod_{x\in\mathcal{X}}\frac{q_{xit}^{\kappa\mu_{xit}-1}(1-q_{xit})^{\kappa(1-\mu_{xit})-1}}{B(\kappa\mu_{xit}, \kappa(1-\mu_{xit}))}\prod_{t=s_i}^{T_i}\frac{p(\boldsymbol{\epsilon}_i^{(t)})}{g(\boldsymbol{\epsilon}_i^{(t)})}\bigg\}.
\end{equation}

The Monte Carlo approximation for the expected values in (\ref{equation:country-marginal}) is computationally expensive, making the computation of, e.g., the WAIC based on the marginal likelihood infeasible since we need to evaluate it for all elements of the posterior sample. Therefore, we use the BIC metric since we only need to evaluate it once. We compute the BIC on the posterior mean of the parameters. The marginal BIC (now denoted by BIC$_m$) for $v$ parameters and $N$ observations is defined as:

$$\text{BIC}_{m} = -2\log p(\mathbf{q}\mid\hat{\Lambda}, \hat{\boldsymbol{\phi}}, \hat{\boldsymbol{\sigma}}) + v\log(N).$$

We also compute the \waic, which is the WAIC based on the conditional likelihood (\ref{equation:cond-likelihood}) and has been used in \cite{Revuelta_Hidalgo_2020} to compare beta factor models. We compute it using the \texttt{R} package \texttt{loo} \citep{Gelman_Hwang_2013,Vehtari_Gelman_2017}.

\subsection{Model fit}

To assess model fit, the root mean squared error (RMSE) and the mean absolute percentage error (MAPE) between observed probabilities of death and reconstructed probabilities of death from posterior predictive distributions are considered. We aim to verify if the model is reconstructing the observed probabilities of death well. The predicted posterior value for the probability of death for age group  $x$, country $i$ at time $t$ is denoted by $\hat{q}_{xit}$, which is the mean of the posterior predictive distribution based on the model. The RMSE and MAPE are defined, respectively, as:

\begin{equation}\label{equation:rmse}
\text{RMSE}(\mathbf{q}) = \sqrt{\sum_{i=1}^{n}\sum_{t=s_i}^{T_i}\sum_{x\in\mathcal{X}}\frac{(\hat{q}_{xit}-q_{xit})^2}{N}}, \:\text{and  } \text{MAPE}(\mathbf{q}) = \frac{100}{N}\sum_{i=1}^{n}\sum_{t=s_i}^{T_i}\sum_{x\in\mathcal{X}}\bigg |\frac{\hat{q}_{xit}-q_{xit}}{q_{xit}}\bigg |.   
\end{equation}

Finally, we compute the Spearman correlation between observed mortality distances and latent effect distances to assess how well the latent effects preserve the order of Euclidean distances between observations—a measure known as cophenetic correlation \citep{Sokal_Rohlf_1962,Kraemer_Reichstein_2018}. For any two country-time pairs $(i,t)$ and $(j,t')$, we define two types of distances:

\begin{equation}\label{equation:distance}
  d_{it, jt'}^{\text{obs}} = \sqrt{\sum_{x\in\mathcal{X}}(q_{xit} - q_{xjt'})^2}, \quad \text{and} \quad d_{it, jt'}^{\text{lat}} = \sqrt{\sum_{k=1}^{K}(\theta_{ik}^{(t)} - \theta_{jk}^{(t')})^2},
\end{equation}
where $d_{it, jt'}^{\text{obs}}$ is the Euclidean distance in the observed mortality space (across all age groups), and $d_{it, jt'}^{\text{lat}}$ is the Euclidean distance in the latent space. Let $\mathbf{D}^{\text{obs}}$ denote the vector of all pairwise observed mortality distances, and $\mathbf{D}^{\text{lat}}$ denote the vector of all pairwise latent effect distances. The cophenetic correlation is then $\text{Corr}(\mathbf{D}^{\text{obs}}, \mathbf{D}^{\text{lat}})$.

Additionally, we compute RMSE and MAPE between the observed distances $\mathbf{D}^{\text{obs}}$ and the distances based on the posterior predictive means $\mathbf{D}^{\text{pred}}$ (where mortality predictions $\hat{q}_{xit}$ replace observed values in the distance calculation), denoted by $\text{RMSE}(\mathbf{D})$ and $\text{MAPE}(\mathbf{D})$, respectively.

To have reference values for the RMSE, MAPE, and the Spearman correlation under the time-dependent BLV model, we also fit a Bayesian Gaussian factor analysis model (denoted BFA here) with the same dimension $K$, which is a widely used model for dimension reduction, and with results comparable to principal component analysis (PCA). We fit the model to the centred $\text{logit}(q_{xit})$ as in \cite{Murphy_Viroli_2020} for a fixed $K$.

\section{Results}\label{section:results}

\subsection{Simulation study}\label{section:simulation-study}

A simulation study was conducted to evaluate the performance of the HMC estimation procedure, the Procrustean transform, and the model-selection criteria (\bic and \waic). To construct realistic simulation scenarios, we first fitted the time-dependent BLV model to the observed data using HMC, and then fixed the generating model's parameters to the estimated values (see Section \ref{section:hmd-application}). Based on these parameters, we generated fifty replicate data sets under two configurations, with true $K = 2$ and $K = 4$.

For each replicate data set, we fitted models with $K = 1,\dots,6$. All time-dependent BLV models were estimated using HMC-NUTS with a chain size of 4000 iterations, including 2000 warm-up iterations, and three parallel chains. The \bic was approximated using the marginal likelihood with a Monte Carlo sample size of 100000. Convergence was assessed using the $\hat{R}$ and effective sample size diagnostics \citep{Gelman_Carlin_2013}, with further details provided in Appendix \ref{appendix:sim-conve-check}. The code for the simulation study is available at \href{https://github.com/pedroaraujo9/tblv}{github.com/pedroaraujo9/tblv}. 

\begin{figure}[H]
\centering
\includegraphics[scale=0.55]{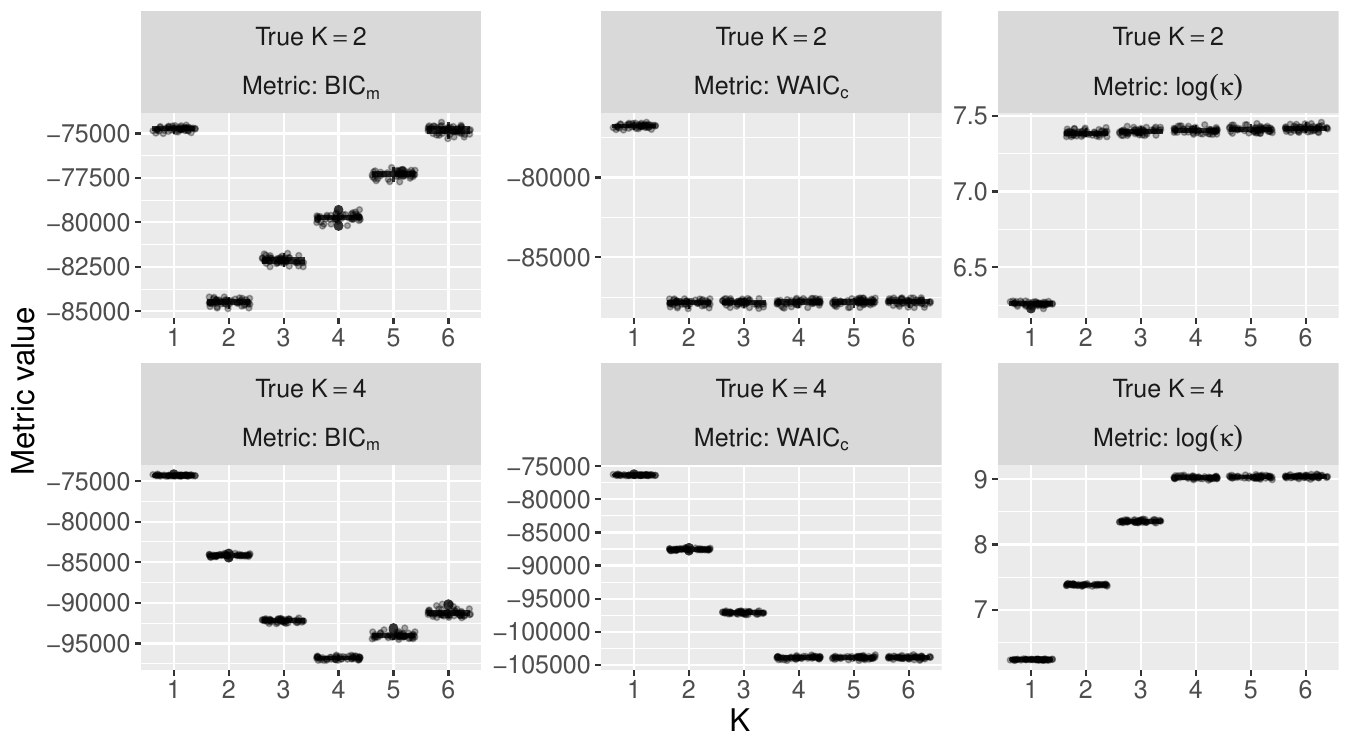}
\caption{Boxplots of \bic, \waic and $\log(\kappa)$ for each replicate and true $K \in \{2, 4\}$.}
\label{fig:sim-study:sim-model-selection}
\end{figure}

Table \ref{table:sim:metric} details the distribution of the selected $K$ under the \bic and \waic metrics. The \waic tends to overestimate the number of latent effects in some cases, while the \bic performs better, always selecting the correct number of dimensions. Additionally, Figure \ref{fig:sim-study:sim-model-selection} shows the boxplot of the value for each of the metrics for each replicate and the posterior mean of $\log(\kappa)$ for different $K$. The \waic and $\log(\kappa)$ have an elbow at the true $K$, and the \bic always has its lowest value at the true $K$.

\begin{longtable}{lcccc}
\caption{Distribution of the selected \( K \) for the metrics \bic and \waic when the true \( K \) is 2 or 4.}
\label{table:sim:metric} \\

\toprule
& \multicolumn{2}{c}{True \( K = 2 \)} & \multicolumn{2}{c}{True \( K = 4 \)} \\
\cline{2-3}\cline{4-5}
Selected \( K \) & \bic & \waic & \bic & \waic \\
\midrule
\endfirsthead

\toprule
& \multicolumn{2}{c}{True \( K = 2 \)} & \multicolumn{2}{c}{True \( K = 4 \)} \\
\cline{2-3}\cline{4-5}
Selected \( K \) & \bic & \waic & \bic & \waic \\
\midrule
\endhead

\midrule
\multicolumn{5}{r}{\textit{Continued on next page}}
\endfoot

\bottomrule
\endlastfoot

1   & \z   & \z   & \z   & \z   \\
2   & 50   & 41   & \z   & \z   \\
3   & \z   & 9    & \z   & \z   \\
4   & \z   & \z   & 50   & 36  \\
5   & \z   & \z   & \z   & 9   \\
6   & \z   & \z   & \z   & 5   \\

\end{longtable}

\begin{figure}[H]
  \centering
  \begin{tabular}{@{}c@{}} 
    \includegraphics[width=0.7\textwidth]{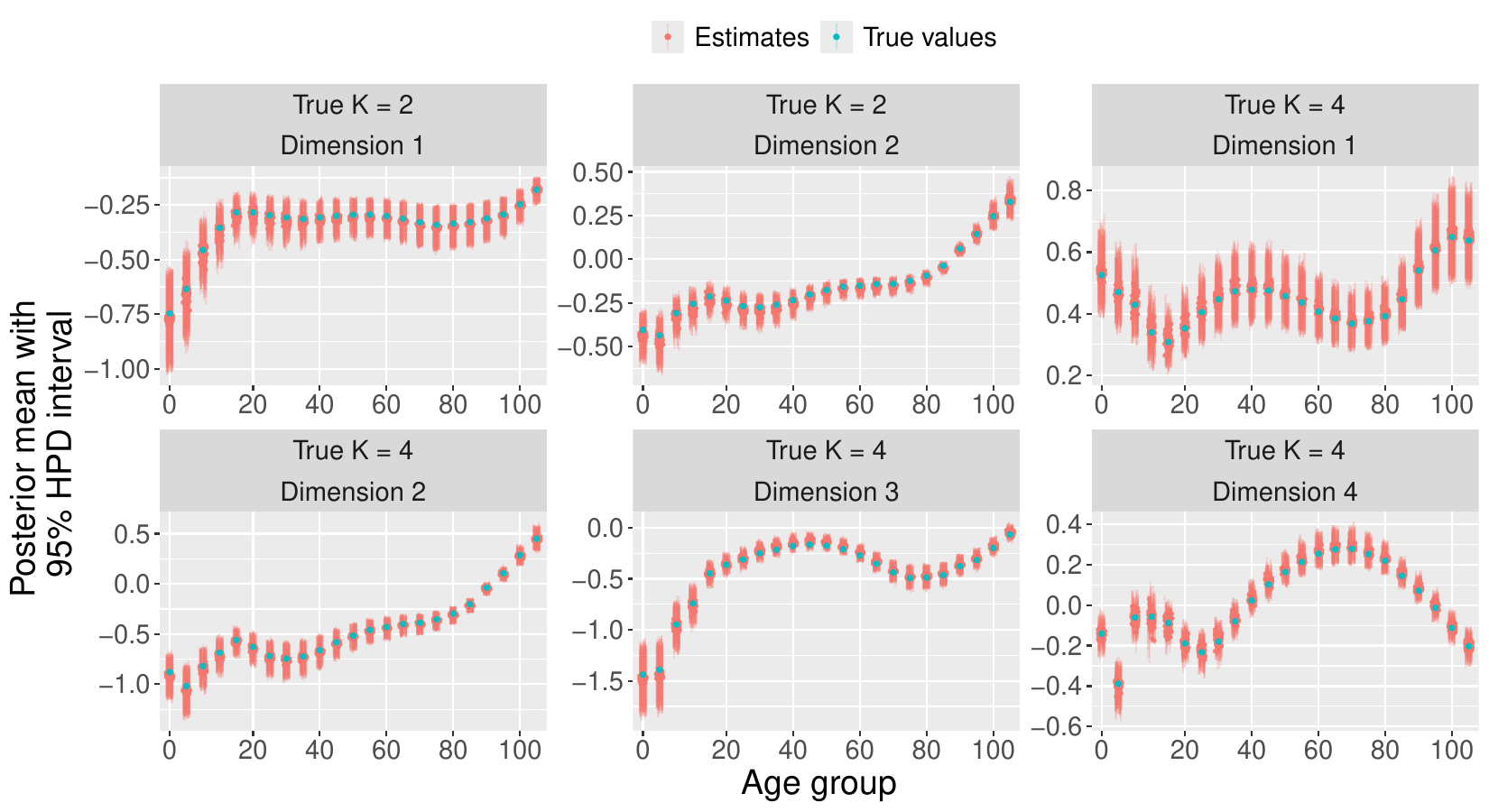} \\
    \small (a) Estimated values with 95\% HPD for all replicates of $\alpha_{xk}$. \\[\floatsep]
    \includegraphics[width=0.7\textwidth]{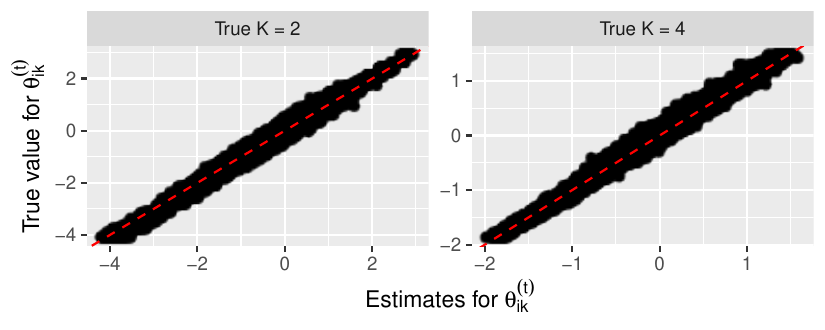} \\
    \small (b) Estimated values for $\theta_{ik}^{(t)}$ for all replicates when $K$ is fixed at its true value of 2 or 4.
  \end{tabular}
  \caption{Estimated values with 95\% HPD for $\alpha_{xk}$ in (a) and $\theta_{ik}^{(t)}$ in (b) for all replicates when $K$ is fixed at its true value of 2 or 4.}
  \label{fig:sim-study:alpha-theta}
\end{figure}

Figure \ref{fig:sim-study:alpha-theta} shows the estimates for $\alpha_{xk}$ and $\theta_{ik}^{(t)}$ when $K$ is fixed at its true value. To address identifiability, the estimated values of $\alpha_{xk}$ and $\theta_{ik}^{(t)}$ were rotated with a Procrustean transform to best match the true values.
We observe that the estimation is accurate, with the true and estimated values matching in most cases, with more variability for $K=2$.

\begin{figure}[H]
\centering
\includegraphics[scale=0.4]{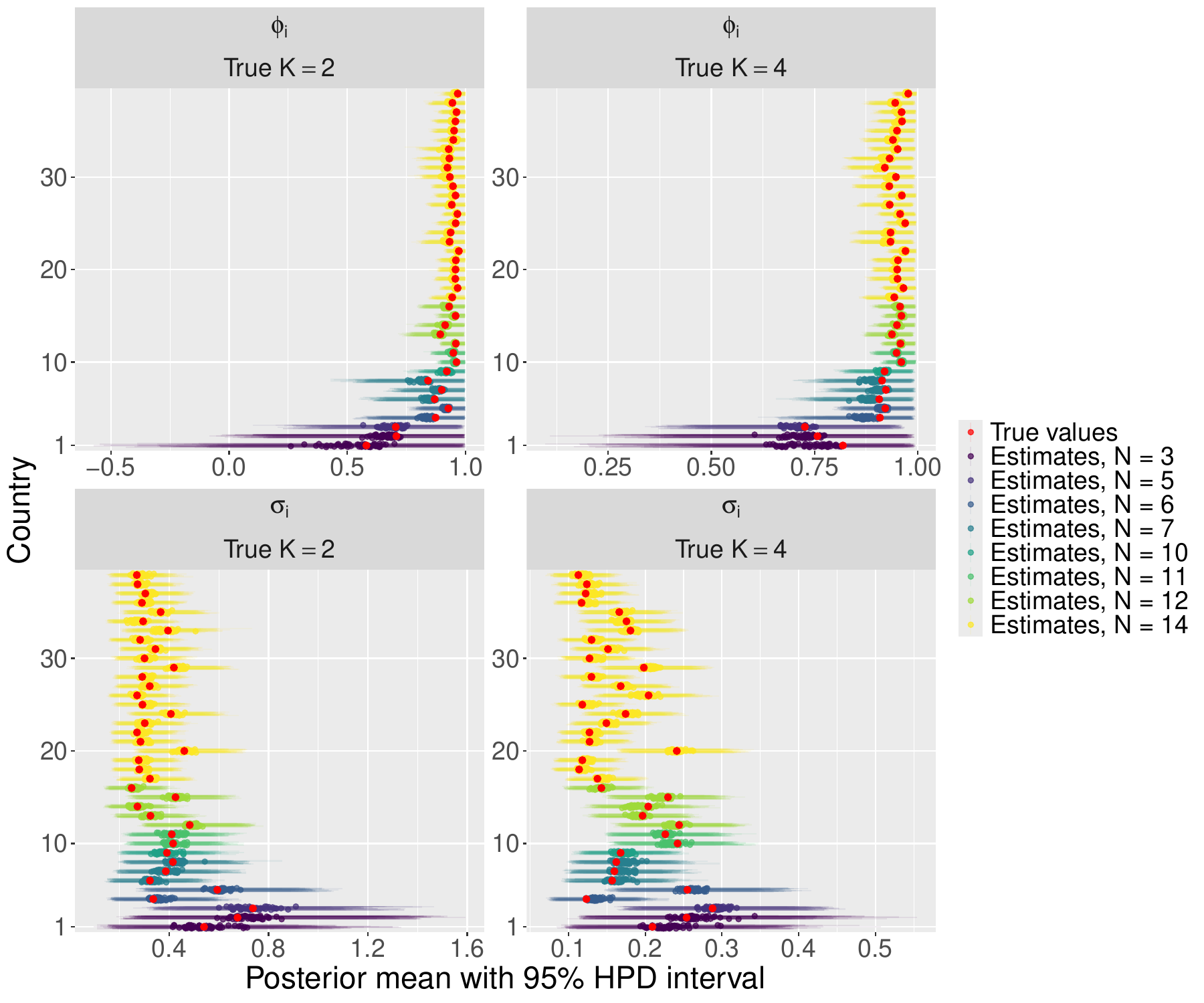}
    \caption{Estimated values, 95\% HPD and true values (in red) for $\phi_{i}$ and $\sigma_{i}$ for $K\in\{2, 4\}$ fixed at its true value. 
    Countries are coloured by their number of available time points, $N$, 
    going from 3 (purple) to 14 (yellow).
    }
    \label{fig:sim-study:sigma-phi}
\end{figure}

Figure \ref{fig:sim-study:sigma-phi} shows the estimated value for the 50 replicates and true values for $\phi_{i}$ and for $\sigma_{i}$ along with the 95\% HPDs. For $\phi_{i}$, the parameter is well estimated when the number of time points is not small. The parameter $\phi_{i}$ seems to be underestimated in cases where the number of time points available is low, and the uncertainty for those cases is big. For $\sigma_{i}$, we see that the overall estimation is good, and again countries with lower data will present more uncertainty.

\subsection{HMD application}\label{section:hmd-application}

We apply the time-dependent BLV model with $K=1,2,\dots, 10$ to the HMD data described in Section \ref{section:data}, generating samples from the posterior distribution with the NUTS setup from \texttt{Stan} with standard hyperparameters and using 3 chains for the parameters in all models. The chain size was 15000, with a warm-up of 10000 and a thinning size of 10, giving a final posterior sample of size 1500. The marginal likelihood was approximated as outlined in Section \ref{section:model-choice} with a Monte Carlo sample size of 200000. Convergence of the chains was assessed by analysing $\hat{R}$, and the effective number of parameters \citep{Gelman_Carlin_2013}; further details are provided in Appendix \ref{appendix:conve-check}. 

\subsubsection{Model choice}\label{section:model-choice}

\begin{figure}[H]
\centering
\includegraphics[width=0.8\textwidth]{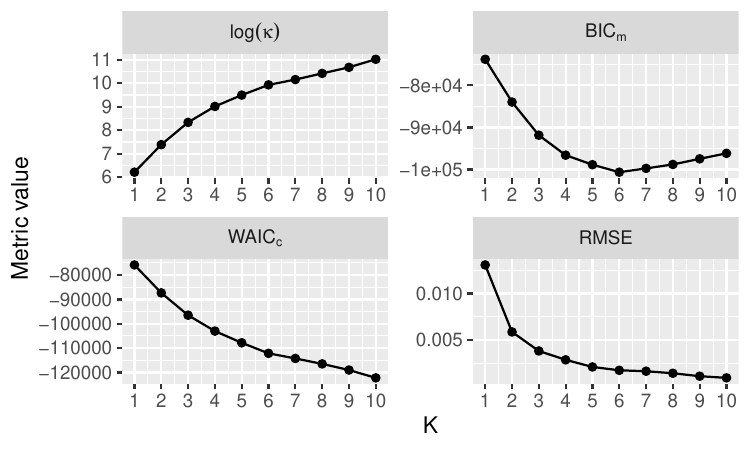}
\caption{Model choice metrics for the time-dependent BLV model with $K=1,\dots, 10$.}
\label{fig:analysis-metrics}
\end{figure}

Figure \ref{fig:analysis-metrics} shows the $\log(\kappa)$, \bic, RMSE and \waic for the HMD data for each $K=1,\dots, 10$. In the simulation study, the \bic demonstrated good ability when selecting the true dimension; here, $K=6$ has the lowest value for this metric. The RMSE drops rapidly from $K = 1$ to $K=6$, with little relative change thereafter. The $\log(\kappa)$ and \waic metrics are inconclusive in terms of the optimal $K$. The model with $K=6$ is desirable as it offers a good reconstruction of the observed mortalities with fewer and more interpretable parameters. Therefore, the model with $K=6$ was selected for further analysis since it presents the lowest \bic and provides a parsimonious and interpretable model. The $\log(\kappa)$ for this model is 9.92, with 95\% HPD interval $[9.89, 9.95]$.

\subsubsection{Parameter estimates}\label{section:model-param}

\begin{figure}[H]
\centering
\includegraphics[scale=0.6]{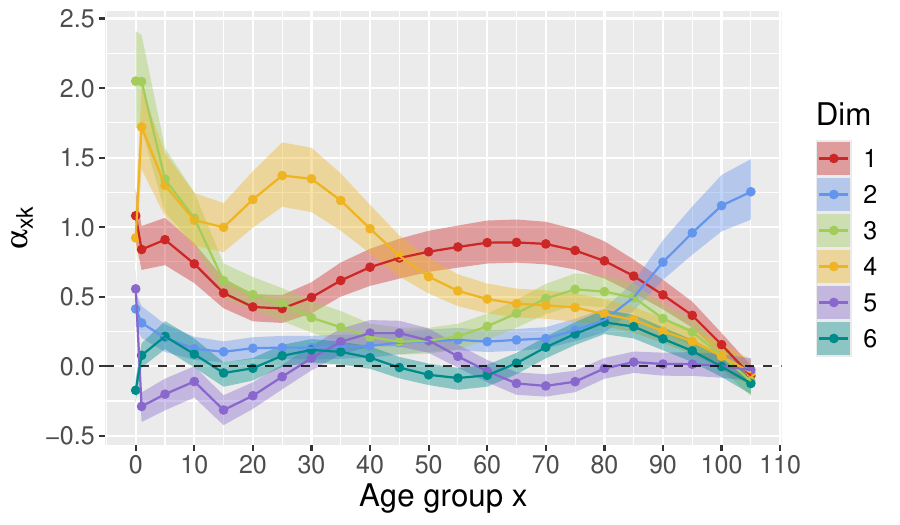}
\caption{Posterior mean and 95\% HPD interval for $\alpha_{xk}$.}\label{fig:results:alpha-beta}
\end{figure}

To increase the interpretability of $\alpha_{xk}$, we applied the varimax transformation to the posterior mean of $\alpha_{xk}$ and rotated the posterior samples to match this solution. Figure \ref{fig:results:alpha-beta} shows the posterior mean of $\alpha_{xk}$ with its 95\% HPD intervals. We see that most latent effects affect newborn mortality. In addition to this, the first dimension captures overall mortality, with more weight on young and middle age groups; the second dimension captures mortality in very old age groups; dimension 3 captures child mortality and some elderly mortality; dimension 4 captures child and young-adult mortality; whilst dimensions 5 and 6 capture some extra variability.

\begin{figure}[H]
\centering
\includegraphics[width=0.98\textwidth]{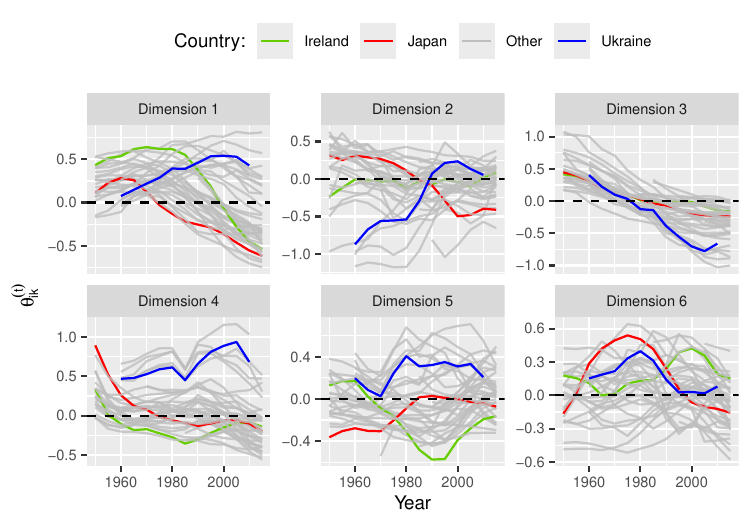}
\caption{Posterior mean $\theta_{ik}^{(t)}$ for each latent dimension over time, with some countries highlighted. Dimension 1 is ``overall mortality", dimension 2 is ``elderly mortality," dimension 3 is ``young mortality," and dimension 4 is ``young-adult mortality." Remaining dimensions capture variability with no clear pattern. The black dashed line represents zero effect.}
\label{fig:results:theta-est}
\end{figure}

Figure \ref{fig:results:theta-est} shows the latent effects over time in each dimension, with some countries highlighted. For dimension 1 (overall mortality), countries such as Ukraine experienced a substantial increase in the 1990s before declining in recent years, while both Japan and Ireland show decreasing trends, with Japan declining more rapidly. Dimension 2 (elderly mortality) reveals a similar pattern: Ukraine exhibited an increase in the 1980s, Ireland remained relatively stable, and Japan lowered its effect. Dimension 3 (young mortality) has declined for almost all countries, consistent with the trends observed in Figure \ref{fig:data:tau}. For dimension 4 (young-adult mortality), a clear division emerges between developed countries such as Japan and Ireland and Eastern European countries such as Ukraine; the remaining effects capture additional variability with no clear pattern. Overall, the latent effects successfully capture the temporal evolution of mortality in a manner consistent with the observed data.

\begin{figure}[H]
\centering
\includegraphics[width=0.75\textwidth]{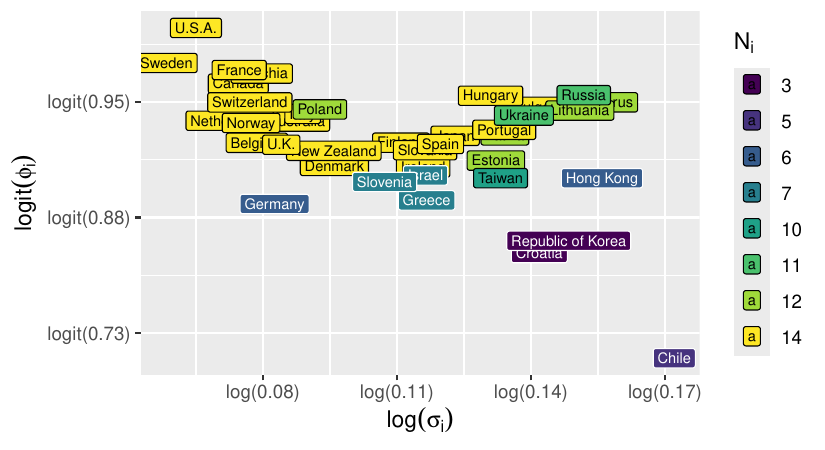}
\caption{Posterior means of $\phi_{i}$ and $\sigma_{i}$ for each country $i$ under the time-dependent BLV model with $K=6$.}
\label{fig:results:sigma-phi-est}
\end{figure}

Figure \ref{fig:results:sigma-phi-est} shows the autocorrelation $\phi_{i}$ on the logit scale and the standard deviation $\sigma_{i}$ on the log scale for each country coloured by the number of time points $N_i$. In general, large, developed countries have high values of $\phi_i$ and lower $\sigma_{i}$, indicating a high temporal association and lower volatility in the latent effect. Smaller developed countries such as Spain, Ireland, and Finland have similar estimates of $\phi_{i}$ and $\sigma_{i}$, with weaker autocorrelation and larger variance. Countries such as Russia, Ukraine, Belarus, Poland, and the Baltic countries also have high values of $\phi_i$ along with a large variance, which indicates higher variability in the mortality trajectory of those countries over time. Small countries and countries with less data, such as Chile and Hong Kong, have lower $\phi_{i}$ and increased $\sigma_{i}$ estimates, with intuitively wider 95\% HPD intervals.

\subsubsection{Model fit}

Table \ref{table:model-fit} shows the RMSE and MAPE for the reconstructed mortality based on the time-dependent BLV for the selected model with $K=6$, and simpler models with $K=2$ and $K=4$ to compare the results based on the posterior predictive distributions of those models. For reference, RMSE and MAPE values under the Bayesian factor analysis model (BFA) applied to the centred $\text{logit}(q_{xit})$, which also selected 6 factors, are also detailed. 

We see that for data matrix $\mathbf{q}$, the time-dependent BLV has a lower RMSE than the BFA model in all dimensions. The MAPE is initially higher for dimension $K=2$ but approaches the BFA values for $K=4$ and $K=6$. For the distance matrix $\mathbf{D}$, the BLV is better than the BFA in most scenarios, indicating the time-dependent BLV can better preserve the distances between observations on the original mortality scale. When considering the latent effect distances, the BLV is consistently better than the BFA.

\begin{table}[H]
\caption{The RMSE and MAPE for the data matrix $\mathbf{q}$ and Euclidean distances $\mathbf{D}$. The RMSE for $\mathbf{q}$ and $\mathbf{D}$ is multiplied by 100 to increase the scale. $\text{Corr}(\mathbf{D}^{\text{obs}}, \mathbf{D}^{\text{lat}})$ is the Spearman correlation (cophenetic correlation) between the observed mortality distances and the latent effect distances. The time-dependent beta latent variable model is denoted BLV; BFA denotes the Bayesian Gaussian factor analysis model applied to the logit probability of death.\label{table:model-fit}}
\begin{tabular*}{\columnwidth}{@{\extracolsep{\fill}}lcccccc@{\extracolsep{\fill}}}
\toprule
Model & $K$ & $\text{RMSE}(\mathbf{q})$ & $\text{MAPE}(\mathbf{q})$ & RMSE$(\mathbf{D})$ & MAPE$(\mathbf{D})$ &  $\text{Corr}(\mathbf{D}^{\text{obs}}, \mathbf{D}^{\text{lat}})$ \\
\midrule
BLV & 2 & 0.588 & 14.454 & 1.61  & 8.403 & 0.832 \\
BFA & 2 & 1.34  & 10.565 & 5.32  & 20.379 & 0.649 \\
\midrule
BLV & 4 & 0.288 & 6.322  & 0.555 & 3.131 & 0.566 \\
BFA & 4 & 0.363 & 6.801  & 0.873 & 4.388 & 0.474 \\
\midrule
BLV & 6 & 0.176 & 4.125  & 0.173 & 1.066 & 0.512 \\
BFA & 6 & 0.252 & 4.016  & 0.613 & 2.592 & 0.329 \\
\bottomrule
\end{tabular*}
\end{table}

\begin{figure}[H]
\centering
\includegraphics[scale=0.8]{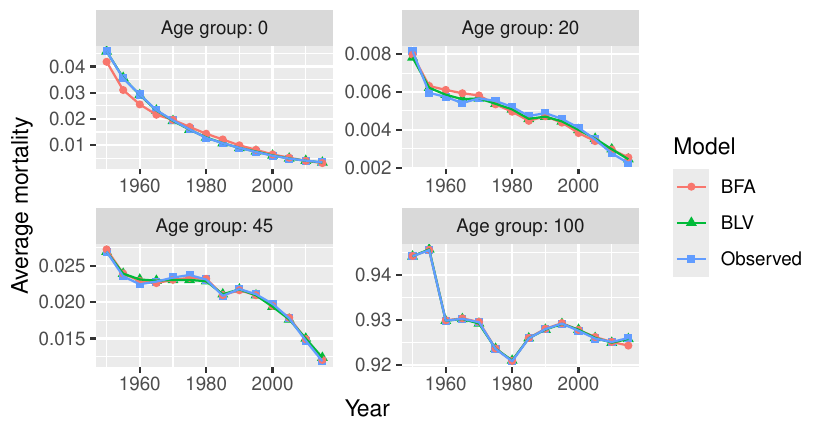}
\caption{Posterior mean predicted mortality for some age groups over time for the time-dependent beta latent variable model (BLV) and for the Bayesian Gaussian factor analysis model (BFA) model (applied to the $\text{logit}$ mortality), both with $K=6$. }
\label{fig:results:post-pred-check}
\end{figure}

Figure \ref{fig:results:post-pred-check} shows, for some age groups, the posterior-predicted mean probability of death under both models with $K=6$, alongside the observed values. We see, for example, that the BFA fits slightly worse for the age group $[0, 1)$, while the BLV captures mortality in that age group well, with the lines closely following the observed data. Overall, the time-dependent BLV with $K=6$ fits well, reconstructing the observed mortality.

\section{Conclusion and discussion}\label{section:conclusion}

We have seen that country-level probabilities of death over time form a complex multivariate dataset, relevant to both governments and private companies. The analysed mortality data lie in the interval (0, 1) and vary over time with different dynamics and associations. We proposed a time-dependent beta latent variable (BLV) model for analysing mortality data. Our goal was to summarise the mortality profile with a few time-dependent latent effects. The time-dependent BLV model does not require transforming the data and accounts for country-specific time dependence. This model provides an interesting alternative to the standard approach of modelling logit-transformed mortality with a normal distribution. We estimated the parameters under the Bayesian paradigm, using Hamiltonian Monte Carlo (HMC) to generate samples from the posterior distribution. The best latent dimension $K$ was selected using the BIC, based on the marginal likelihood with respect to the latent effects.

Our simulation study showed that the HMC-NUTS estimation procedure works well for this class of models, consistent with findings in \cite{Revuelta_Hidalgo_2020}. Model selection based on the \bic approximation also performed well. Notably, the fact that the \bic based on the marginal log-likelihood outperformed the \waic based on the conditional log-likelihood supports the perspective from \cite{Merkle_Edgar_2019} that marginal likelihood-based metrics are preferable for latent variable models. However, since our simulation was tailored to our specific application, further research is needed to assess the performance of our estimation procedure and metrics in other scenarios.

When applied to the HMD data, the time-dependent BLV model was able to summarise the observed mortality with $K=6$ latent effects, indicating that mortality dynamics are complex and cannot be summarised with just one or two latent effects. The model provided a straightforward way to visualise countries' mortality over time, and the latent effects can be useful for other tasks. These latent effects allowed us to observe interesting behaviours in the mortality profile without having to analyse the mortality data for each of the 23 age groups separately. The estimates for $\phi_{i}$ and $\sigma_{i}$ were also informative and, when visualised together, provided a good summary of the overall mortality dynamics. We also observed groups of countries with similar behaviour.

Overall, the time-dependent BLV model fits well when compared with the BFA applied to logit-transformed probability of death, suggesting that using the beta distribution to model untransformed data can be more suitable than the Gaussian approach to logit-transformed probability of death, especially for clustering, as the BLV model preserves distances between observations very well.

Several potential areas for future work exist. The model can be generalised to other problems and applied to other datasets of probabilities or rates varying over time. Based on the time-dependent BLV estimates for the HMD countries, it is likely that groups of countries with similar mortality trends exist. An extension of the time-dependent BLV model that clusters country effects and parameters in a model-based manner could reveal such a structure, enhancing our understanding of global mortality dynamics, following works such as \cite{Murphy_Viroli_2020} and \cite{Leger_Mazzuco_2021}. 

We also observed large changes in country effects over time; a changepoint model for the latent effects \citep{Liu_Zhang_2022} could be useful for detecting them automatically. Unit-interval data can also contain values of 0 and 1, and extending the model to accommodate zero and one inflation, as in \cite{Molenaar_Curi_2022}, may be necessary. In addition to model innovations, finding more efficient ways to estimate the parameters is crucial because the estimation time for the HMC algorithm was long (an average of 3 hours for one fit on a computing cluster); variational methods could be an alternative to reduce computational time \citep{Loaiza-Maya_Smith_2022}. Variational approaches are generally much faster than MCMC, though they would require additional work, such as developing a reliable and efficient optimisation procedure. A full comparison is beyond the scope of this paper, but future work could examine whether this yields meaningful computational gains without compromising estimation quality.

In summary, the time-dependent BLV model with a single precision parameter is a good alternative for modelling mortality data when the goal is to find patterns in the data, as it provides useful lower-dimensional representations of the dataset. The time-dependent BLV model parsimoniously represents observed mortality data through a few sets of country latent effects that are easier to analyse. The latent effects generated by the model can be a valuable input for further analysis, such as clustering, changepoint detection, or even forecasting.

\begin{appendices}

\section{Data details}\label{appendix:data-availability}

\begin{figure}[H]
\centering
\includegraphics[width=0.8\textwidth]{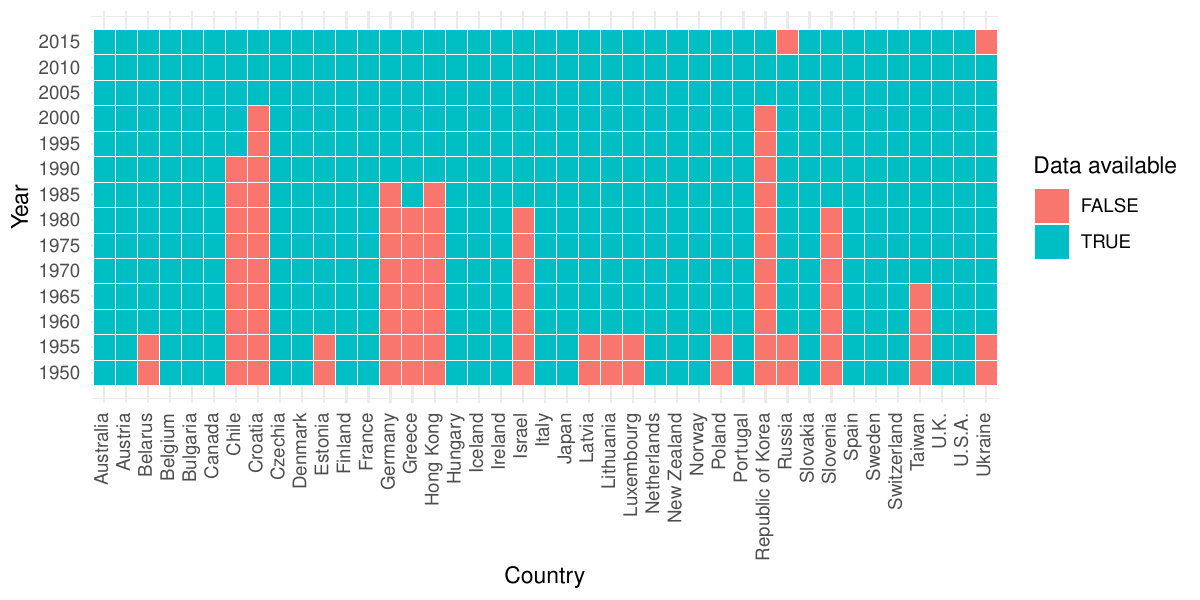}
\caption*{Data availability for different countries over time.}
\label{fig:data:availability}
\end{figure}

\section{Convergence metrics for simulation study}\label{appendix:sim-conve-check}

\begin{figure}[H]
\centering
\includegraphics[width=0.8\textwidth]{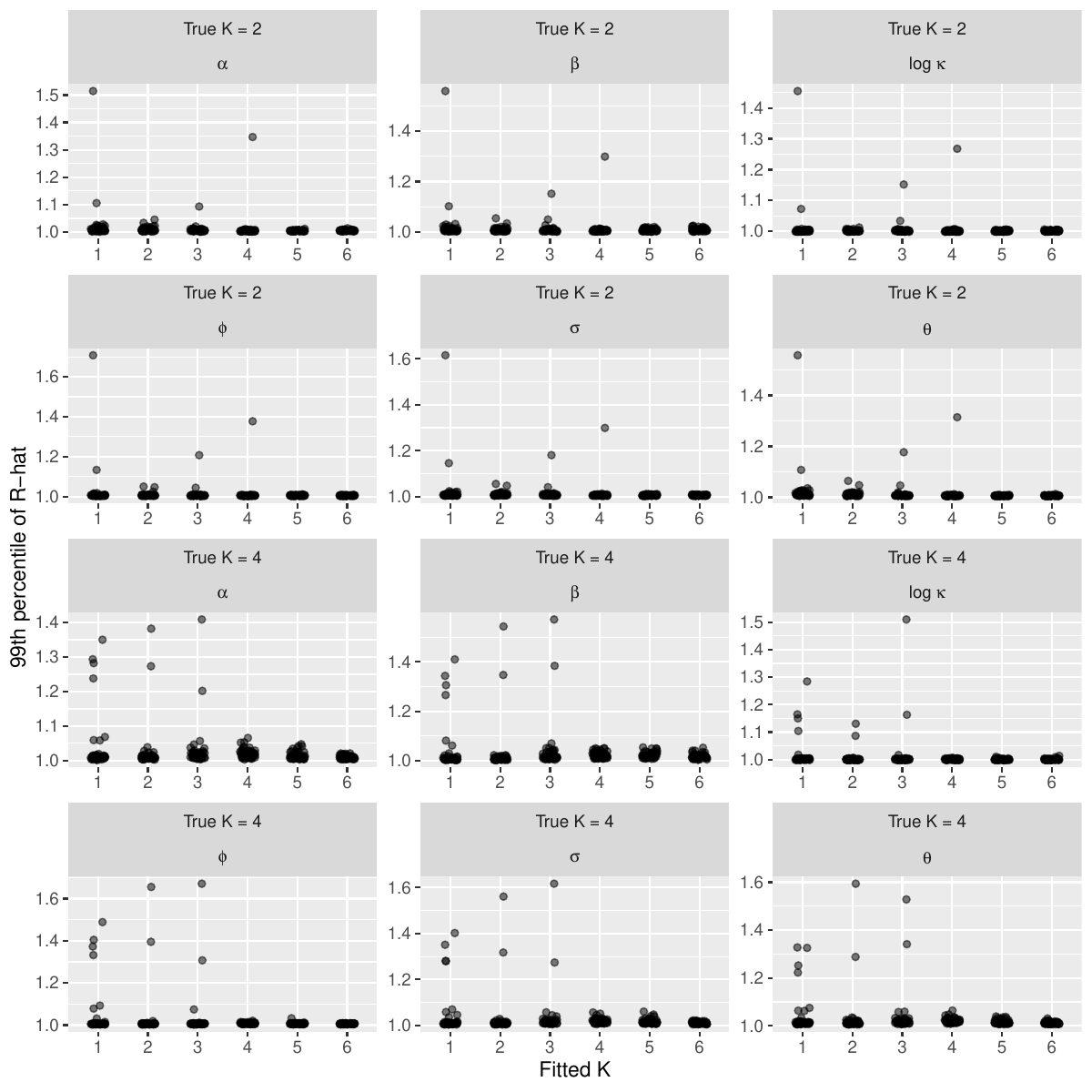}
\caption*{The 99th percentile of $\hat{R}$ for each group of parameters and each $K$ when the true $K$ is 2 or 4.}
\label{fig:sim-study:convergence-check-rhat}
\end{figure}

\begin{figure}[H]
\centering
\includegraphics[width=0.8\textwidth]{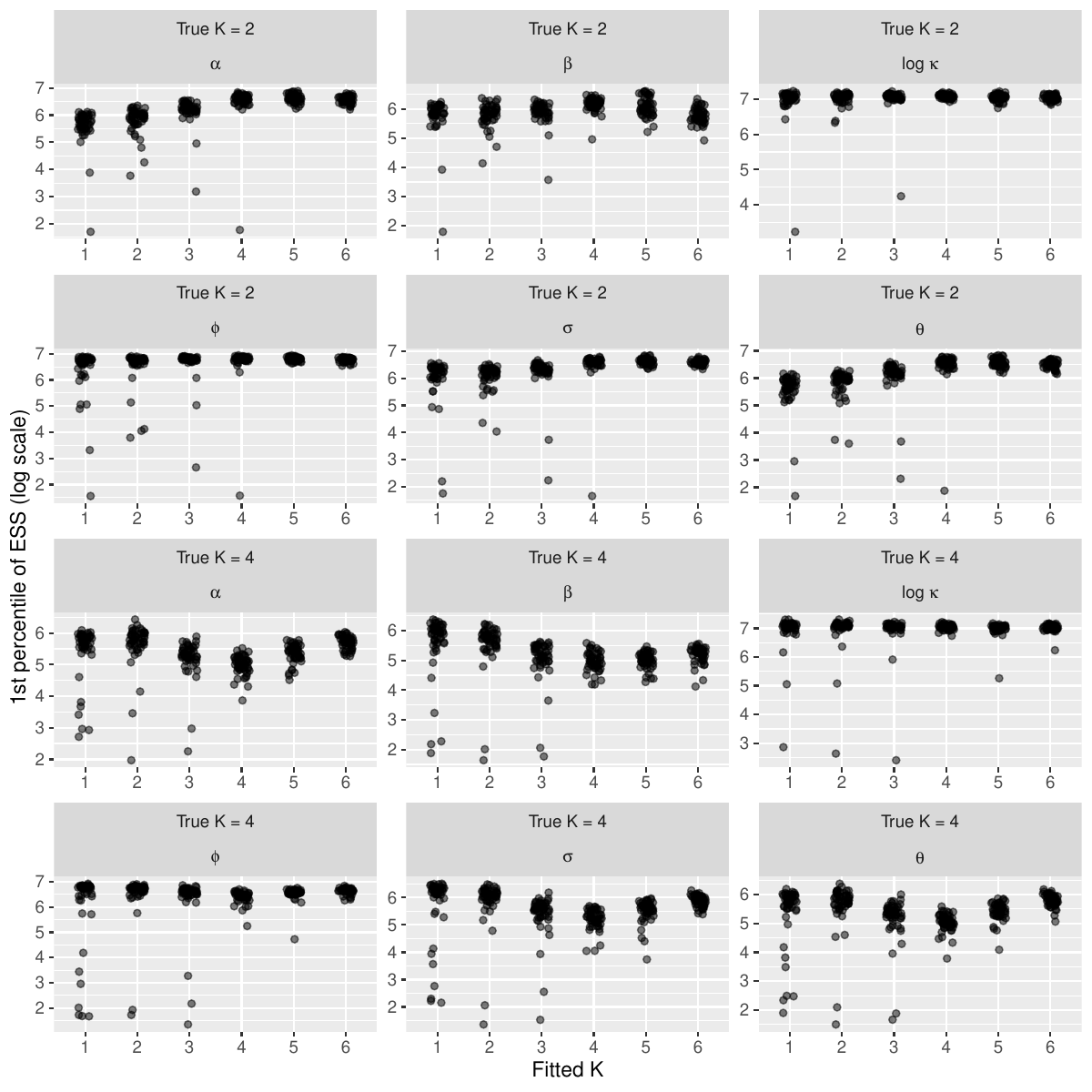}
\caption*{1st percentile of effective sample size for each group of parameters and each $K$ when the true $K$ is 2 or 4.}

\label{fig:sim-study:convergence-check-ess}
\end{figure}

\section{Convergence metrics for the HMD application}\label{appendix:conve-check}

\begin{table}[H]
\caption{Proportion of times $\hat{R} > 1.1$ for each group of parameters. \label{table:prop-R-greater-1.1}}
\begin{tabular*}{\columnwidth}{@{\extracolsep{\fill}}ccccccc@{\extracolsep{\fill}}}
\toprule
$K$ & $\alpha$ & $\beta$ & $\log(\kappa)$ & $\log(\sigma)$ & $\phi$ & $\theta$ \\
\midrule
1 & 0.00 & 0.00 & 0.00 & 0.00 & 0.00 & 0.00 \\ 
2 & 0.00 & 0.00 & 0.00 & 0.00 & 0.00 & 0.00 \\ 
3 & 0.00 & 0.00 & 0.00 & 0.00 & 0.00 & 0.00 \\ 
4 & 0.00 & 0.00 & 0.00 & 0.00 & 0.00 & 0.00 \\ 
5 & 0.00 & 0.00 & 0.00 & 0.00 & 0.00 & 0.00 \\ 
6 & 0.00 & 0.00 & 0.00 & 0.00 & 0.00 & 0.00 \\ 
7 & 0.00 & 0.00 & 0.00 & 0.00 & 0.00 & 0.00 \\ 
8 & 0.00 & 0.00 & 0.00 & 0.00 & 0.00 & 0.00 \\ 
9 & 0.00 & 0.00 & 0.00 & 0.00 & 0.00 & 0.00 \\ 
10 & 0.00 & 0.00 & 0.00 & 0.00 & 0.00 & 0.00 \\
\bottomrule
\end{tabular*}
\end{table}

\begin{table}[H]
\caption{Proportion of times the effective sample size is lower 10 for each group of parameters. \label{table:prop-effective-sample-size}}
\begin{tabular*}{\columnwidth}{@{\extracolsep{\fill}}ccccccc@{\extracolsep{\fill}}}
\toprule
$K$ & $\alpha$ & $\beta$ & $\log(\kappa)$ & $\log(\sigma)$ & $\phi$ & $\theta$ \\
\midrule
1 & 0.00 & 0.00 & 0.00 & 0.00 & 0.00 & 0.00 \\ 
2 & 0.00 & 0.00 & 0.00 & 0.00 & 0.00 & 0.00 \\ 
3 & 0.00 & 0.00 & 0.00 & 0.00 & 0.00 & 0.00 \\ 
4 & 0.00 & 0.00 & 0.00 & 0.00 & 0.00 & 0.00 \\ 
5 & 0.00 & 0.00 & 0.00 & 0.00 & 0.00 & 0.00 \\ 
6 & 0.00 & 0.00 & 0.00 & 0.00 & 0.00 & 0.00 \\ 
7 & 0.00 & 0.00 & 0.00 & 0.00 & 0.00 & 0.00 \\ 
8 & 0.00 & 0.00 & 0.00 & 0.00 & 0.00 & 0.00 \\ 
9 & 0.00 & 0.00 & 0.00 & 0.00 & 0.00 & 0.00 \\ 
10 & 0.00 & 0.00 & 0.00 & 0.00 & 0.00 & 0.00 \\
\bottomrule
\end{tabular*}
\end{table}

\end{appendices}

\section{Acknowledgments}

This publication has emanated from research conducted with the financial support of Taighde Éireann – Research Ireland under Grant number 18/CRT/6049. For the purpose of Open Access, the author has applied a CC BY public copyright licence to any Author Accepted Manuscript version arising from this submission.

\bibliographystyle{abbrvnat}
\bibliography{reference}

\end{document}